*Article*

# Water Dynamics in Highly Concentrated Protein Systems—Insight from Nuclear Magnetic Resonance Relaxometry

Danuta Kruk [1,*], Adam Kasparek [1], Elzbieta Masiewicz [1], Karol Kolodziejski [1], Radosław Cybulski [2] and Bartosz Nowak [2]

[1] Department of Physics and Biophysics, University of Warmia & Mazury in Olsztyn, Oczapowskiego 4, 10-719 Olsztyn, Poland; adam.kasparek@uwm.edu.pl (A.K.); elzbieta.masiewicz@uwm.edu.pl (E.M.); karol.kolodziejski@uwm.edu.pl (K.K.)
[2] Department of Mathematical Methods of Informatics, University of Warmia & Mazury in Olsztyn, Sloneczna 54 Street, 10-710 Olsztyn, Poland; radoslaw.cybulski@uwm.edu.pl (R.C.); bnowak@matman.uwm.edu.pl (B.N.)
* Correspondence: danuta.kruk@uwm.edu.pl

**Abstract:** [1]H spin-lattice relaxation experiments have been performed for water–Bovine Serum Albumin (BSA) mixtures, including 20%wt and 40%wt of BSA. The experiments have been carried out in a frequency range encompassing three orders of magnitude, from 10 kHz to 10 MHz, versus temperature. The relaxation data have been thoroughly analyzed in terms of several relaxation models with the purpose of revealing the mechanisms of water motion. For this purpose, four relaxation models have been used: the data have been decomposed into relaxation contributions expressed in terms of Lorentzian spectral densities, then three-dimensional translation diffusion has been assumed, next two-dimensional surface diffusion has been considered, and eventually, a model of surface diffusion mediated by acts of adsorption to the surface has been employed. In this way, it has been demonstrated that the last concept is the most plausible. Parameters describing the dynamics in a quantitative manner have been determined and discussed.

## 1. Introduction

The dynamical properties of molecular systems are one of the most fundamental questions of molecular science. The question encompasses not only the time scale of the motion but also its mechanism—in other words: one is not satisfied with determining the time scale of a specific dynamical process (by providing, for instance, diffusion coefficients), one wishes to get insight into the geometry of the motion (for instance the dimensionality of the translation displacements).

The experimental means allowing to enquire into the characteristic features of molecular motion are very limited. Nuclear Magnetic Resonance (NMR) methods are broadly appreciated as a source of information about molecular structure and dynamics. As far as dynamics are concerned, NMR relaxation studies are of primary importance. However, "classical" NMR relaxation experiments are commonly performed at a single, high magnetic field (resonance frequency). According to spin relaxation theories [1–3], the relaxation process is most efficient when the time scale of the fluctuations of the spin interactions causing the relaxation is of the order of the reciprocal resonance frequency. This implies that at high frequencies, one mostly probes fast dynamics. Consequently, to probe dynamical processes occurring over a broad time scale, one has to vary the magnetic field (resonance frequency). This kind of study is referred to as NMR relaxometry. In the present studies, the resonance frequency is varied from about 10 kHz to 10 MHz ($^1$H resonance frequency), which gives three orders of magnitude. Consequently, one can probe molecular motion on the time scale from about $10^{-4}$ s to about $10^{-8}$ s in a single experiment. This potential of NMR relaxometry has been widely exploited for molecular and ionic systems of varying complexity—from liquids [4–6] via polymers and proteins [7–20] to tissues [21,22] and liquid and solid electrolytes [23–30]. The great advantage of NMR relaxometry is the ability to give insight into the mechanism of motion. Relaxation rates (reciprocal relaxation times) are given as linear combinations of so-called spectral density functions. A spectral density function is defined as a Fourier transform of a corresponding time correlation function characterizing the stochastic fluctuations (caused by the molecular motion) of the spin interactions. The mathematical form of the correlation function (and, hence, the spectral density) depends on the mechanism of the motion (for instance, such as isotropic and anisotropic rotational motion, free (three-dimensional) translation diffusion, or restricted (two-dimensional or one-dimensional) translation motion). Via the form of the spectral density function, the shape of the frequency dependence of the relaxation rates is a fingerprint of the mechanism of motion. At this stage, one should point out that single-frequency relaxation studies hardly contain information about the characteristic features (mechanisms) of the dynamical process leading to the relaxation at that frequency.

The advantages of NMR relaxometry (the ability to probe molecular motion over a broad time scale and the ability to reveal the mechanism of motion) interfere with each other. The reason for that is several relaxation contributions present over such a broad frequency range and constitute the overall relaxation rate. The relaxation contributions stem from different relaxation pathways. For instance, magnetic dipole-dipole interactions (being the dominating origin of $^1$H relaxation) can be of intra-molecular or inter-molecular origin. The first ones fluctuate in time as a result of rotational and internal dynamics, while the second ones are mostly modulated by translation diffusion. For simple systems, the two contributions can be unambiguously identified and disentangled [5,6], profiting from the time scale separation of translational and rotational dynamics. In such a case, one can fully profit from the unique advantages of NMR relaxometry (investigating rotational and translational dynamics in a single experiment and identifying the mechanism of the observed dynamical processes) in a relatively straightforward way. The task becomes much more cumbersome for multi-component systems due to several relaxation contributions and not clear time scale separation of the dynamical processes associated with the relaxation contribution. Examples of such systems are highly concentrated protein—water mixtures. The systems include a macromolecular fraction (proteins) forming a matrix entrapping water molecules. Consequently, one can expect pools of water molecules to perform different kinds of complex motions.

The purpose of this work is twofold. The first one is to enquire into the mechanism of water motion in the presence of a substantial fraction of proteins (in contrast to highly diluted protein solutions [18]), profiting from the unique potential of NMR relaxometry. For this purpose, Bovine Serum Albumin (BSA) has been chosen as an example. In this context, one should mention NMR relaxometry studies for sedimented proteins showing much different dynamics than proteins in solution [19] and studies addressing the subject of water diffusion on protein surfaces in the presence of ions [20]. The second goal has methodological aspects. We present a thorough analysis of $^1$H spin-lattice relaxation data for BSA–water mixtures using different forms of spectral density functions. In this way, we demonstrate the challenges of revealing the mechanisms of molecular motion for complex systems. At the same time, the work presents an overview of theoretical models that can potentially be exploited to reproduce NMR relaxometry data for systems including water and a macromolecular fraction and illustrates by examples their verification.

*Theory*

$^1$H NMR spin-lattice relaxation processes are predominantly caused by magnetic dipole-dipole interactions. According to the spin relaxation theory [1–3], the spin-lattice relaxation rate, $R_1$, originating from $^1$H-$^1$H dipole-dipole interactions, is given as the following combination of spectral density functions:

$$R_1(\omega) = C_{DD}[J(\omega) + 4J(2\omega)], \quad (1)$$

where $\omega$ denotes the resonance frequency in angular frequency units, while $C_{DD}$ is referred to as a dipolar relaxation constant reflecting the amplitude of the magnetic dipole-dipole interactions causing the relaxation process. The form of the spectral density function, $J(\omega)$ (Fourier transform of the corresponding time correlation function), depends on the mechanism of the motion responsible for stochastic time fluctuations of the dipole-dipole interactions. For exponential correlation functions, the Fourier transform (and, hence, the spectral density) takes a Lorentzian form. Consequently, the relaxation rate is given as [1–3]:

$$R_1(\omega) = C_{DD}\left[\frac{\tau_c}{1+(\omega\tau_c)^2} + \frac{4\tau_c}{1+(2\omega\tau_c)^2}\right], \quad (2)$$

where $\tau_c$ denotes a time constant characterizing the time scale of the motion, referred to as a correlation time. As already pointed out in the Introduction, the broad frequency range covered in NMR relaxometry experiments implies that several dynamical processes can be probed in a single experiment. The simplest way to get some insight into the molecular motion is to attempt to decompose the overall relaxation process into contributions associated with dynamics occurring on different timescales. In such a case, the relaxation rate can be expressed as [12,16,17]:

$$R_1(\omega) = C_s^{DD}\left[\frac{\tau_s}{1+(\omega\tau_s)^2} + \frac{4\tau_s}{1+(2\omega\tau_s)^2}\right] + C_i^{DD}\left[\frac{\tau_i}{1+(\omega\tau_i)^2} + \frac{4\tau_i}{1+(2\omega\tau_i)^2}\right] + C_f^{DD}\left[\frac{\tau_f}{1+(\omega\tau_f)^2} + \frac{4\tau_f}{1+(2\omega\tau_f)^2}\right] + A, \quad (3)$$

where $\tau_s, \tau_i$, and $\tau_f$ denote correlation times characterizing slow, intermediate, and fast dynamics (in a relative scale), respectively, while $C_s^{DD}, C_i^{DD}$, and $C_f^{DD}$ are the corresponding dipolar relaxation constants. The frequency-independent factor, $A$, accounts for a relaxation contribution associated with a very fast motion for which the condition: $\omega\tau_c \ll 1$ is fulfilled in the whole frequency range. An example of such dynamics can be the movement of water molecules in bulk. The decomposition assumes that the contributing dynamical processes can be characterized by exponential correlation functions.

One can go beyond the simple description (parametrization) and attempt to get insight into the mechanism of the molecular motion. In water—protein mixtures, it is expected that water molecules perform translation diffusion that is considerably affected by the presence of the macromolecules. Discussing translation diffusion, one should consider the dimensionality of this process—the translation motion can be isotropic (three-dimensional) or anisotropic (two-dimensional in this case). The two-dimensional translation diffusion one envisages a motion occurring near the surface of the macromolecules (surface diffusion). The spectral density function for three-dimensional diffusion, $J_{3D}(\omega)$, takes the form [31–33]:

$$J_{3D}(\omega) = \frac{72}{5}\int_0^\infty \frac{u^4}{81+9u^2-2u^4+u^6}\frac{\tau_{trans}}{1+(\omega\tau_{trans})^2}du, \quad (4)$$

Consequently, the corresponding expression for the spin-lattice relaxation rate, $R_1(\omega)$, can be expressed as a sum of a relaxation contribution associated with three-dimensional translation diffusion and Lorentzian terms. Limiting ourselves to a single Lorentzian term, one obtains:

$$R_1(\omega) = \frac{3}{2}\left(\frac{\mu_0}{4\pi}\gamma_H^2\hbar\right)^2 \frac{1}{d^3} N_H \int_0^\infty \frac{u^4}{81+9u^2-2u^4+u^6} \left[\frac{\tau_{trans}}{u^4+(\omega\tau_{trans})^2} + \frac{4\tau_{trans}}{u^4+(2\omega\tau_{trans})^2}\right] du + C_{DD}\left[\frac{\tau_c}{1+(\omega\tau_c)^2} + \frac{4\tau_c}{1+(2\omega\tau_c)^2}\right] + A, \quad (5)$$

where $\gamma_H$ is $^1$H gyromagnetic factor, $\mu_0$ is the vacuum permeability, $\hbar$ is reduced Planck constant, $N_H$ denotes the number of hydrogen atoms per unit volume (referring to the fraction of water molecules undergoing the translation diffusion), while $d$ denotes a distance of the closest approach [31,32]. The model is called force free hard sphere model—it assumes that molecules have a form of hard spheres with $^1$H nuclei placed in their centers. In this approximation, the distance of the closest approach is given as a sum of the radii of the interacting molecules—in case of identical molecules, this gives the molecular diameter. The correlation time $\tau_{trans}$ is given as $\tau_{trans} = \frac{d^2}{2D_{trans}}$, where $D_{trans}$ denotes the translation diffusion coefficient. In the low-frequency range, when $\omega\tau_{trans} < 1$, the spectral density for three-dimensional translation diffusion (Equation (4)) shows a linear dependence on $\sqrt{\omega}$ [31–33]. Consequently, when the dominating relaxation contribution at low frequencies stems from intermolecular dipole-dipole interactions modulated by three-dimensional translation diffusion, the relaxation rates, $R_1(\omega)$, show a linear dependence on $\sqrt{\omega}$ in this range.

In case the diffusion process is restricted to two dimensions—in other words, it occurs in the vicinity of a surface, the corresponding spectral density, $J_{2D}(\omega)$, takes the form [4,15,19,34,35]:

$$J_{2D}(\omega) = \tau_{trans} \ln\left[\frac{1+(\omega\tau_{trans})^2}{\left(\frac{\tau_{trans}}{\tau_{res}}\right)^2+(\omega\tau_{trans})^2}\right], \quad (6)$$

where $\tau_{res}$ denotes a residence lifetime of water molecules on the surface of the macromolecules. For a long residence lifetime, when $\frac{\tau_{trans}}{\tau_{res}} \ll \omega\tau_{trans}$, Equation (6) converges to:

$$J_{2D}(\omega) = \tau_{trans} \ln[1 + (\omega\tau_{trans})^{-2}], \quad (7)$$

This implies that at low frequencies, when $\omega\tau_{trans} < 1$, the spectral density shows a linear dependence on $\ln\omega$ [35]. Therefore, in analogy to the case of three-dimensional diffusion, when the relaxation contribution associated with translation dynamics dominates in the low-frequency range, the relaxation rate shows a linear dependence on $\ln\omega$. For two-dimensional translation diffusion, the counterpart of Equation (5) takes the form:

$$R_1(\omega) = C_{trans}\tau_{trans}\left[\ln\left[\frac{1+(\omega\tau_{trans})^2}{\left(\frac{\tau_{trans}}{\tau_{res}}\right)^2+(\omega\tau_{trans})^2}\right] + 4\ln\left[\frac{1+(2\omega\tau_{trans})^2}{\left(\frac{\tau_{trans}}{\tau_{res}}\right)^2+(2\omega\tau_{trans})^2}\right]\right] + C_{DD}\left[\frac{\tau_c}{1+(\omega\tau_c)^2} + \frac{4\tau_c}{1+(2\omega\tau_c)^2}\right] + A, \quad (8)$$

where $C_{trans}$ denotes a dipolar relaxation constant. When neglecting the effect of the residence lifetime, Equation (8) converges to:

$$R_1(\omega) = C_{trans}\tau_{trans}[\ln(1 + (\omega\tau_{trans})^{-2}) + 4\ln(1 + (2\omega\tau_{trans})^{-2})] + C_{DD}\left[\frac{\tau_c}{1+(\omega\tau_c)^2} + \frac{4\tau_c}{1+(2\omega\tau_c)^2}\right] + A, \quad (9)$$

In biomolecular systems one can also expect a relaxation contribution originating from $^1$H-$^{14}$N dipole-dipole interactions. $^{14}$N nuclei possess quadrupole moments. This implies that in case of slow molecular dynamics, the energy level structure of $^{14}$N nuclei stems from a superposition of their Zeeman and quadrupole interactions. As the quadrupole coupling is independent of the magnetic field, there are magnetic fields at which the $^1$H resonance frequency matches the transition frequencies of the $^{14}$N nucleus between its energy levels. When the $^1$H and $^{14}$N transition frequencies match, the $^1$H magnetization can be transferred to (taken over by) the $^{14}$N nucleus [16,17,36–43]. This manifests itself as a faster decay of the $^1$H magnetization (a higher relaxation rate) at specific frequencies.

The faster decay leads to a frequency-specific enhancement of the spin-lattice relaxation rate, referred to as Quadrupole Relaxation Enhancement (QRE). The $^1$H-$^{14}$N relaxation contribution, $R_1^{H-N}(\omega)$ can be expressed as [43]:

$$R_1^{H-N}(\omega) = C_{DD}^{HN} \times \begin{bmatrix} \left(\frac{1}{3} + \sin^2\theta \cos^2\phi\right)\left(\frac{\tau_Q}{1+(\omega-\omega_-)^2\tau_Q^2} + \frac{\tau_Q}{1+(\omega+\omega_-)^2\tau_Q^2}\right) + \\ \left(\frac{1}{3} + \sin^2\theta \sin^2\phi\right)\left(\frac{\tau_Q}{1+(\omega-\omega_+)^2\tau_Q^2} + \frac{\tau_Q}{1+(\omega+\omega_+)^2\tau_Q^2}\right) + \\ \left(\frac{1}{3} + \cos^2\theta\right)\left(\frac{\tau_Q}{1+(\omega-\omega_0)^2\tau_Q^2} + \frac{\tau_Q}{1+(\omega+\omega_0)^2\tau_Q^2}\right) \end{bmatrix} \quad (10)$$

where the frequencies $\omega_-, \omega_+$ and $\omega_0$ are defined as: $\frac{\omega_-}{2\pi} = a_Q\left(1-\frac{\eta}{3}\right)$, $\frac{\omega_+}{2\pi} = a_Q\left(1+\frac{\eta}{3}\right)$ and $\omega_0 = \omega_+ - \omega_-$, $a_Q$ denotes the quadrupole coupling constant, while $\eta$ is the asymmetry parameter. The angles $\theta$ and $\phi$ describe the orientation of the principal axis system of the electric field gradient tensor with respect to the $^1$H-$^{14}$N dipole-dipole axis, while the correlation time $\tau_Q$ characterizes time fluctuations of the $^1$H-$^{14}$N dipole-dipole coupling. The dipolar relaxation constant, $C_{DD}^{HN}$, is defined as: $C_{DD}^{HN} = \frac{2}{3}\left(\frac{\mu_0}{4\pi}\frac{\gamma_H\gamma_N\hbar}{r_{HN}^3}\right)^2$, where $r_{HN}$ denotes the $^1$H-$^{14}$N inter-spin distance, while $\gamma_N$ denotes $^{14}$N gyromagnetic factor.

## 2. Results

$^1$H spin-lattice relaxation data for BSA–water mixtures, 20%wt and 40%wt of BSA, versus temperature, are shown in Figure 1a,b.

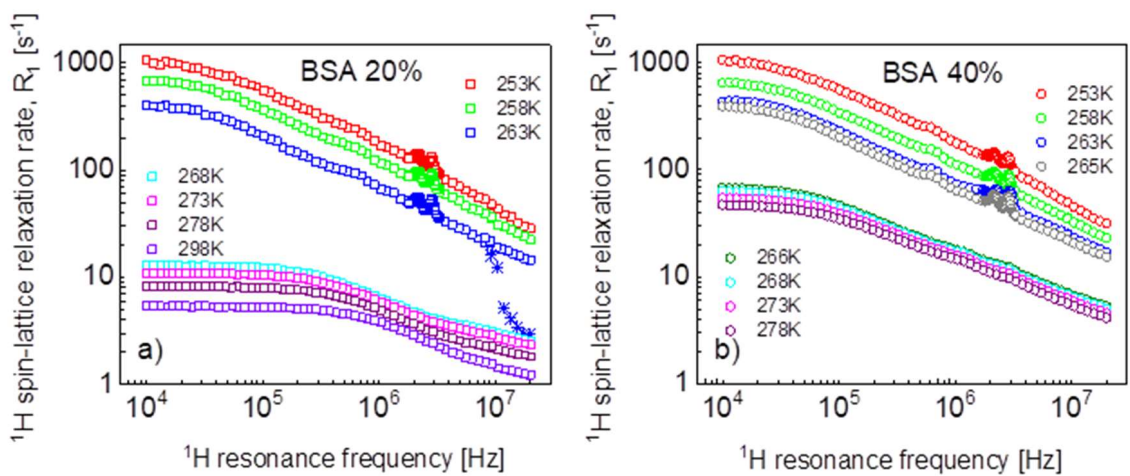

**Figure 1.** $^1$H spin-lattice relaxation rates for BSA–water mixtures versus temperature, (**a**) 20%wt of BSA, (**b**) 40%wt of BSA. Stars show the changes in the relaxation rates upon cooling down.

Looking at Figure 1a, one sees that between 268 K and 263 K, the dynamics of the system changed due to the freezing of the water fraction. Actually, the freezing process has been captured—stars in Figure 1a. The temperature was set to 263 K, and, after 60 min, the experiment began. The relaxation rates at the highest frequency correspond to those at 268 K, then in the course of time with progressing freezing, the relaxation rates reach the values of the relaxation data represented by blue squares that have been obtained at 263 K after waiting the next 60 min. The data for 263 K and below show Quad-rupole Relaxation Enhancement (QRE) effects (quadrupole peaks). For the mixture including 40%wt of BSA (Figure 1b), the freezing temperature has been carefully investigated—it has turned out that at 266 K, the system remains liquid, while it freezes at 265 K. Here, one also sees QRE effects.

Before proceeding with a quantitative analysis of the relaxation data, it is worth noting some effects (Figure 2a). The ratio between the relaxation rates for the mixture containing 40%wt of BSA and 20%wt at 268 K and 273 K has a characteristic shape that, in fact, repeats itself at 278 K (after multiplying the ratio by 0.87). At low temperatures, the ratio reaches a factor close to one in the whole frequency range—that means that the relaxation data tend to overlap. The overlapping is seen in Figure 2b, which also shows, for comparison, relaxation data for solid BSA at 293 K taken from Ref. [16].

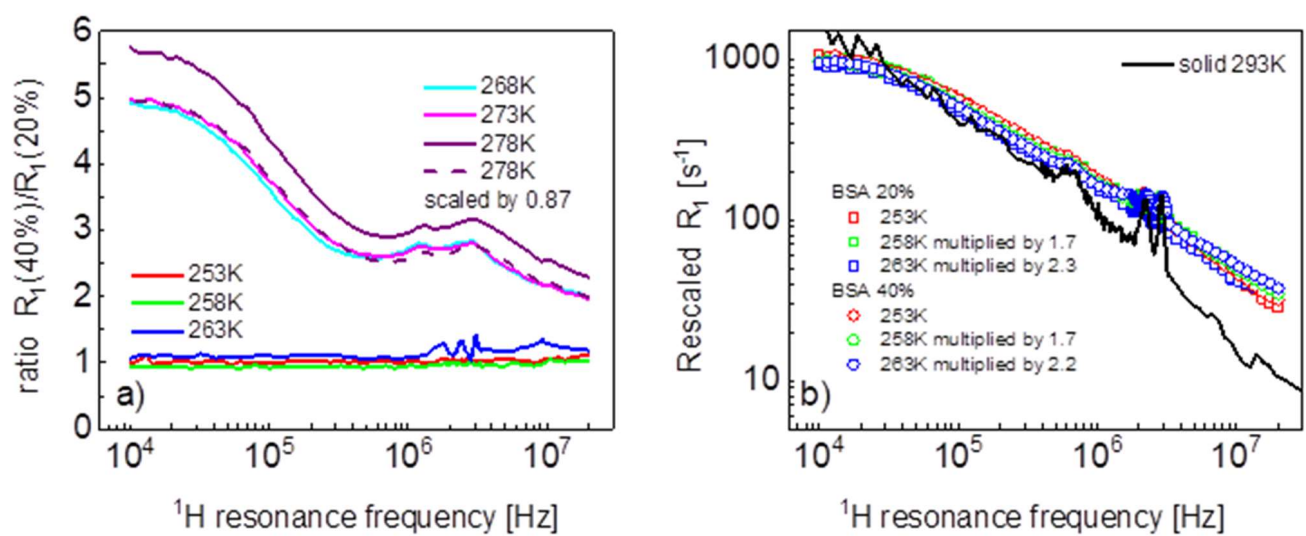

**Figure 2.** (**a**) Ratio between spin-lattice relaxation rates for BSA–water mixtures (40%wt of BSA and 20%wt of BSA); (**b**) rescaled spin-lattice relaxation data for BSA–water mixtures compared with data for solid BSA.

We begin the analysis of the relaxation data with the mixture including 20%wt of BSA and the simplest concept of decomposing the relaxation data into contributions expressed in terms of Lorentzian spectral densities and attributed to dynamical processes referred to as slow, intermediate, and fast ones, according to Equation (2). The outcome of the analysis is shown in Figure 3, while the obtained parameters are collected in Table 1.

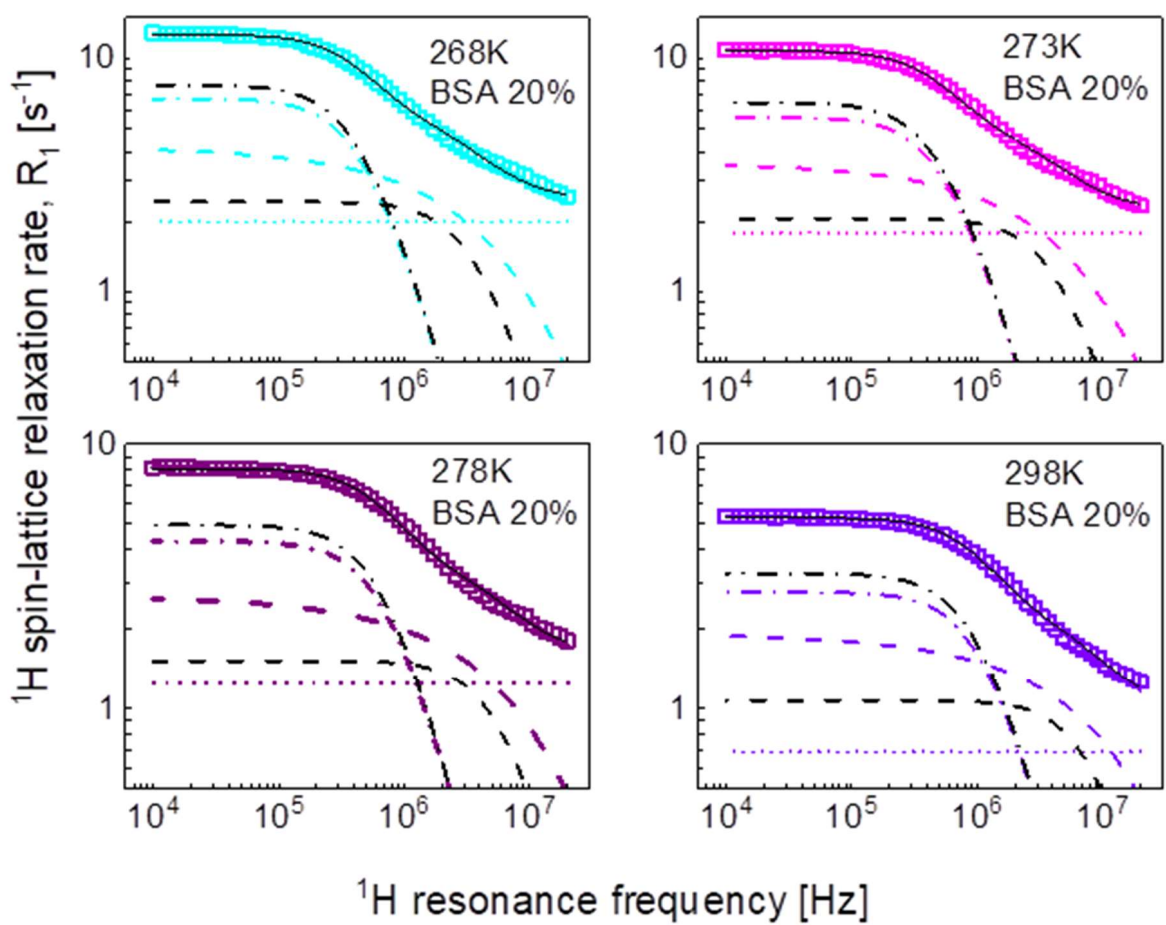

**Figure 3.** $^1$H spin-lattice relaxation data for BSA–water mixture (20%wt of BSA) reproduced in terms of Equation (3); black solid line—overall fit decomposed into a contribution associated with intermediate dynamics (dashed-dotted black line) and fast dynamics (dashed black line), there is no relaxation contribution associated with slow dynamics. Comparison fits obtained in terms of Equation (5) are shown as the corresponding color lines (the lines are hardly visible as they almost overlap with the black ones), they are decomposed into Lorentzian term (dashed-dotted line), a term associated with three-dimensional translation diffusion (dashed line) and frequency independent term (dotted line).

**Table 1.** Parameters obtained from the analysis of $^1$H spin-lattice relaxation data for BSA–water mixtures at higher temperatures (268 K and above) in terms of Equation (1). The dipolar relaxation constants, $C_i^{DD}$ and $C_f^{DD}$, for 20%wt concentration of BSA yields: $C_i^{DD} = 7.84 \times 10^6$ Hz$^2$, $C_f^{DD} = 2.16 \times 10^7$ Hz$^2$. The dipolar relaxation constants for 40%wt concentration of BSA are: $C_s^{DD} = 5.61 \times 10^6$ Hz$^2$, $C_i^{DD} = 1.45 \times 10^7$ Hz$^2$ and $C_f^{DD} = 9.89 \times 10^7$ Hz$^2$.

| 20%wt of BSA | | | |
|---|---|---|---|
| Temp. [K] | $\tau_i$ [s] | $\tau_f$ [s] | $A$ [s$^{-1}$] |
| 268 | $1.97 \times 10^{-7}$ | $2.28 \times 10^{-8}$ | 2.52 |
| 273 | $1.65 \times 10^{-7}$ | $1.91 \times 10^{-8}$ | 2.25 |
| 278 | $1.27 \times 10^{-7}$ | $1.39 \times 10^{-8}$ | 1.62 |
| 298 | $8.33 \times 10^{-8}$ | $9.62 \times 10^{-9}$ | 0.98 |
| 40%wt of BSA | | | |

| Temp. [K] | $\tau_s$ [s] | $\tau_i$ [s] | $\tau_f$ [s] | $A$ [s$^{-1}$] |
|---|---|---|---|---|
| 266 | $1.16 \times 10^{-6}$ | $2.66 \times 10^{-7}$ | $2.24 \times 10^{-8}$ | 4.68 |
| 268 | $1.10 \times 10^{-6}$ | $2.41 \times 10^{-7}$ | $2.09 \times 10^{-8}$ | 4.50 |
| 273 | $9.67 \times 10^{-7}$ | $1.95 \times 10^{-7}$ | $1.84 \times 10^{-8}$ | 3.91 |
| 278 | $8.53 \times 10^{-7}$ | $1.56 \times 10^{-7}$ | $1.59 \times 10^{-8}$ | 3.40 |

For the mixture including 20%wt of BSA, the relaxation data can be reproduced using only two Lorentzian terms (plus the frequency independent term). The obtained parameters have been associated with intermediate and fast dynamics. The association has been made on the basis of the comparison with the parameters obtained for the mixture including 40%wt of BSA. In that case, all three relaxation contributions are needed to reproduce the data, as shown in Figure 4. The order of the values of the longer correlation times obtained for 20%wt of BSA matches that for the correlation times characterizing intermediate dynamics for 40%wt BSA (Table 1). The analysis of the relaxation data for 40%wt of BSA is shown in Figure 4.

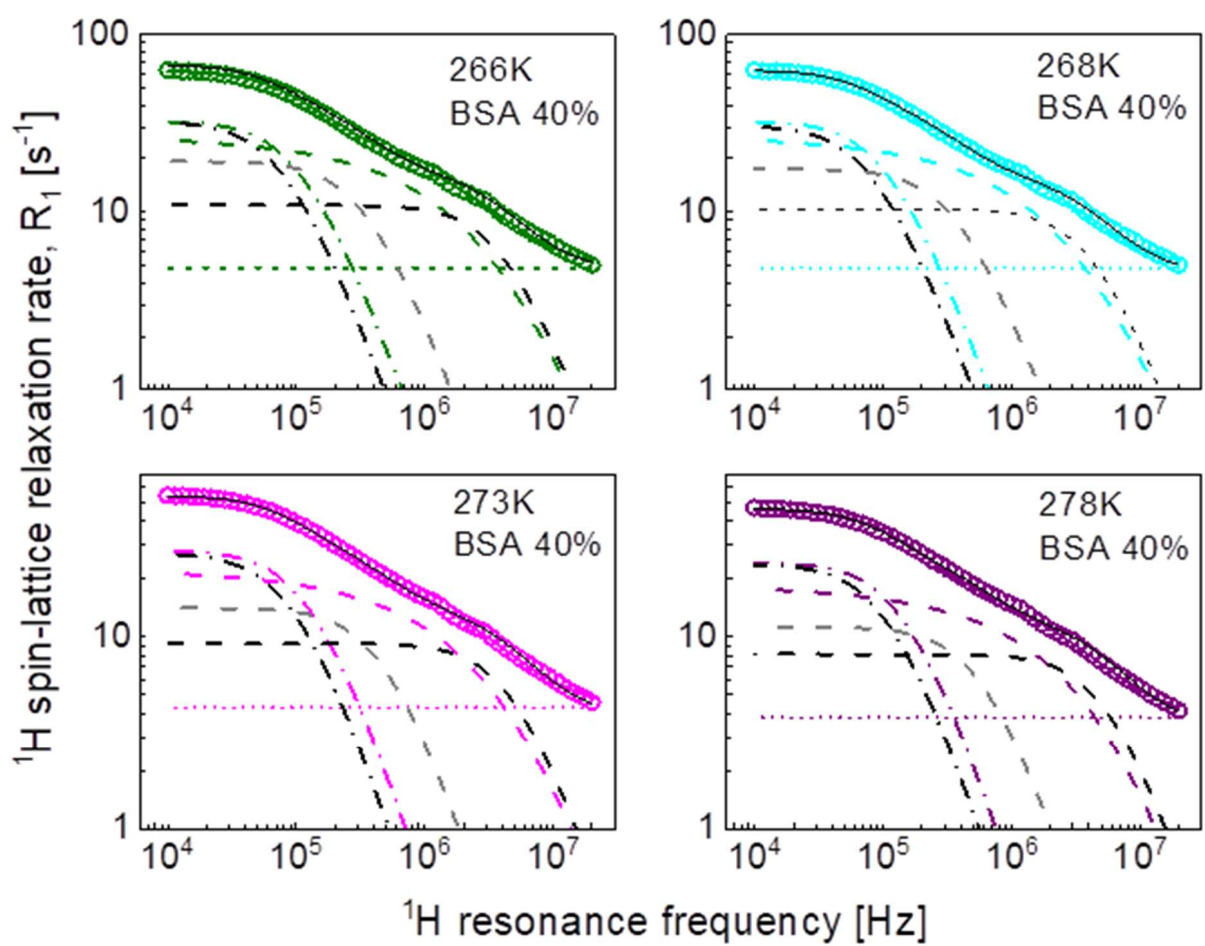

**Figure 4.** $^1$H spin-lattice relaxation data for BSA–water mixture (40%wt of BSA) reproduced in terms of Equation (3); black solid line—overall fit decomposed into a contribution associated with slow dynamics (dashed-dotted black line), intermediate dynamics (dashed grey line), and fast dynamics (dashed black line). Comparison fits obtained in terms of Equation (5) are shown as the corresponding color lines, they are decomposed into a Lorentzian term (dashed-dotted line), a term associated with three-dimensional translation diffusion (dashed line) and a frequency independent term (dotted line).

The obtained parameters give insight into the time scale of the molecular motion, however, we aim at revealing not only the time scale but also the mechanism of the movement of water molecules. Therefore, in the second step we have reproduced the data in terms of Equation (5) as a sum of a relaxation contribution associated with three-dimensional translation diffusion and a Lorentzian term. The fits have been performed with the following adjustable parameters: $C^{DD}, \tau_c, D_{trans}, N_H$ and $A$; the distance of the closest approach has been set to the diameter of a water molecule: $d$ =2.7Å. The parameters obtained for the case of 20%wt of BSA are collated in Table 2. The dipolar relaxation constant, $C^{DD}$=7.29×10$^6$ Hz$^2$ is very close to that obtained for the intermediate dynamics (Equation (3)), $C_i^{DD}$ =7.84×10$^6$ Hz$^2$; the values of the correlation time, $\tau_c$, are also similar to that for $\tau_i$. The fits are shown in Figure 3 for comparison. The same approach has been applied to the relaxation data for the BSA–water (40%wt) mixture. The values of the obtained parameters are discussed in the next section. At this stage one should notice that this approach has led to a reduction in the number of the adjustable parameters—instead of the two pairs of parameters: $C_i^{DD}, \tau_i$ and $C_f^{DD}, \tau_f$, characterizing the intermediate and slow dynamics, the model involves only the translation diffusion coefficient, $D_{trans}$, and the $N_H$ number. The translation diffusion coefficients are rather small.

**Table 2.** Parameters obtained from the analysis of $^1$H spin-lattice relaxation data for BSA–water mixtures in terms of Equation (5). The dipolar relaxation constant, $C_{DD}$, for 20%wt concentration of BSA yields: $C^{DD}$=7.29 × 10$^6$ Hz$^2$, $N_H$ = 1.52 × 10$^{27}$/m$^3$; the corresponding values for 40%wt concentration yield: $C^{DD}$=7.69 × 10$^6$ Hz$^2$, $N_H$ = 2.84 × 10$^{27}$/m$^3$; the distance of closest approach has been set in all cases to $d$ =2.7Å. The correlation time $\tau_{trans}$ has been obtained from the relationship: $\tau_{trans} = \frac{d^2}{2D_{trans}}$.

| Temp. [K] | $\tau_c$ [s] | $D_{trans}$ [m$^2$/s] | $A$ [s$^{-1}$] | $\tau_{trans}$ [s] |
|---|---|---|---|---|
| | | 20%wt of BSA | | |
| 268 | 1.85 × 10$^{-7}$ | 2.08 × 10$^{-12}$ | 1.98 | 3.50 × 10$^{-8}$ |
| 273 | 1.54 × 10$^{-7}$ | 2.44 × 10$^{-12}$ | 1.79 | 2.98 × 10$^{-8}$ |
| 278 | 1.18 × 10$^{-7}$ | 3.28 × 10$^{-12}$ | 1.25 | 2.22 × 10$^{-8}$ |
| 298 | 7.60 × 10$^{-8}$ | 4.56 × 10$^{-12}$ | 0.69 | 1.60 × 10$^{-8}$ |
| | | 40%wt of BSA | | |
| 266 | 9.08 × 10$^{-7}$ | 5.56 × 10$^{-13}$ | 5.00 | 1.31 × 10$^{-7}$ |
| 268 | 8.47 × 10$^{-7}$ | 6.10 × 10$^{-13}$ | 4.84 | 1.20 × 10$^{-7}$ |
| 273 | 7.32 × 10$^{-7}$ | 7.23 × 10$^{-13}$ | 4.31 | 1.01 × 10$^{-7}$ |
| 278 | 6.35 × 10$^{-7}$ | 8.68 × 10$^{-13}$ | 3.82 | 8.40 × 10$^{-8}$ |

In the pursuit of the mechanism of water diffusion, we have attempted to exploit the model of two-dimensional translation diffusion (surface diffusion) represented by Equation (7). The model of two-dimensional translation diffusion combined with a Lorentzian relaxation contribution (Equation (9)) has led to the fits shown in Figure 5 for 20%wt of BSA and in Figure 6 for 40%wt of BSA. The obtained parameters are collated in Table 3.

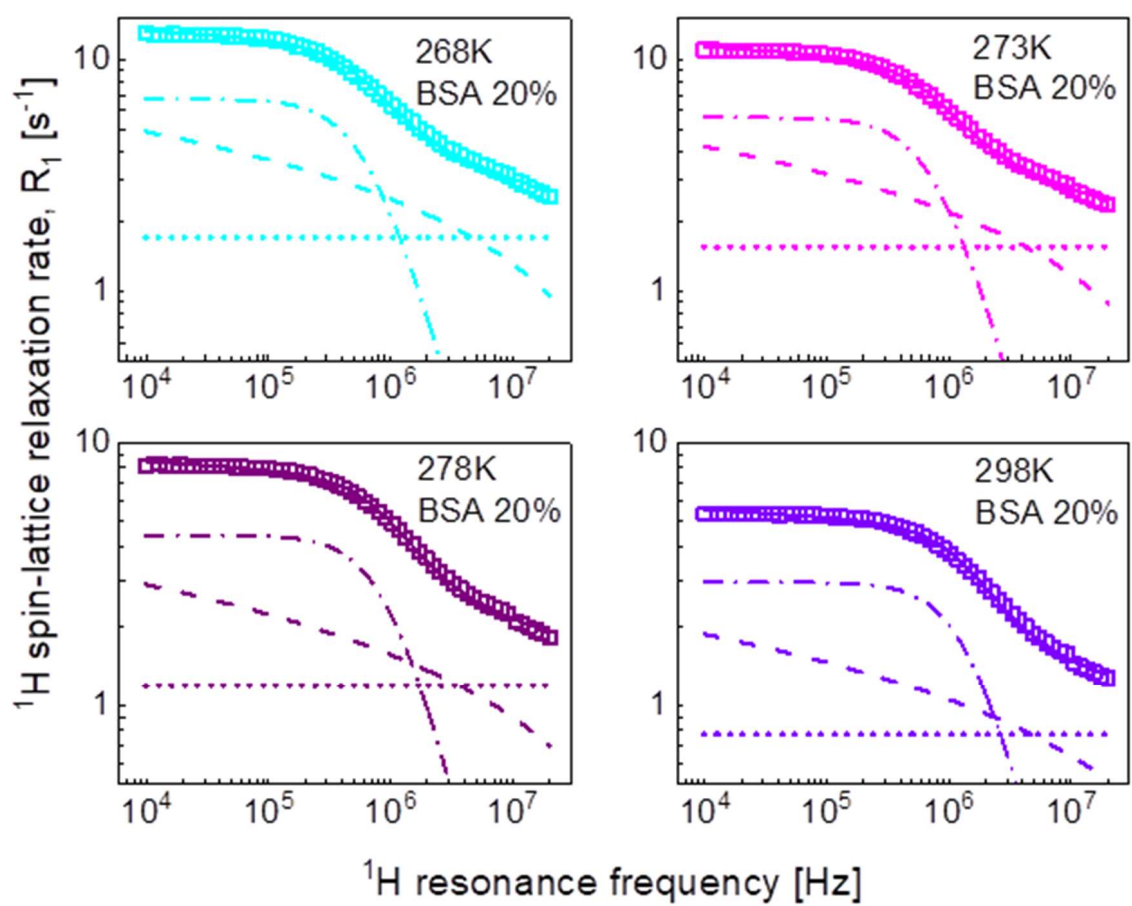

**Figure 5.** $^1$H spin-lattice relaxation data for BSA–water mixture (20%wt of BSA) reproduced in terms of Equation (8); solid line—overall fit decomposed into a Lorentzian term (dashed-dotted line), a term associated with two-dimensional translation diffusion (dashed line) and a frequency independent term (dotted line).

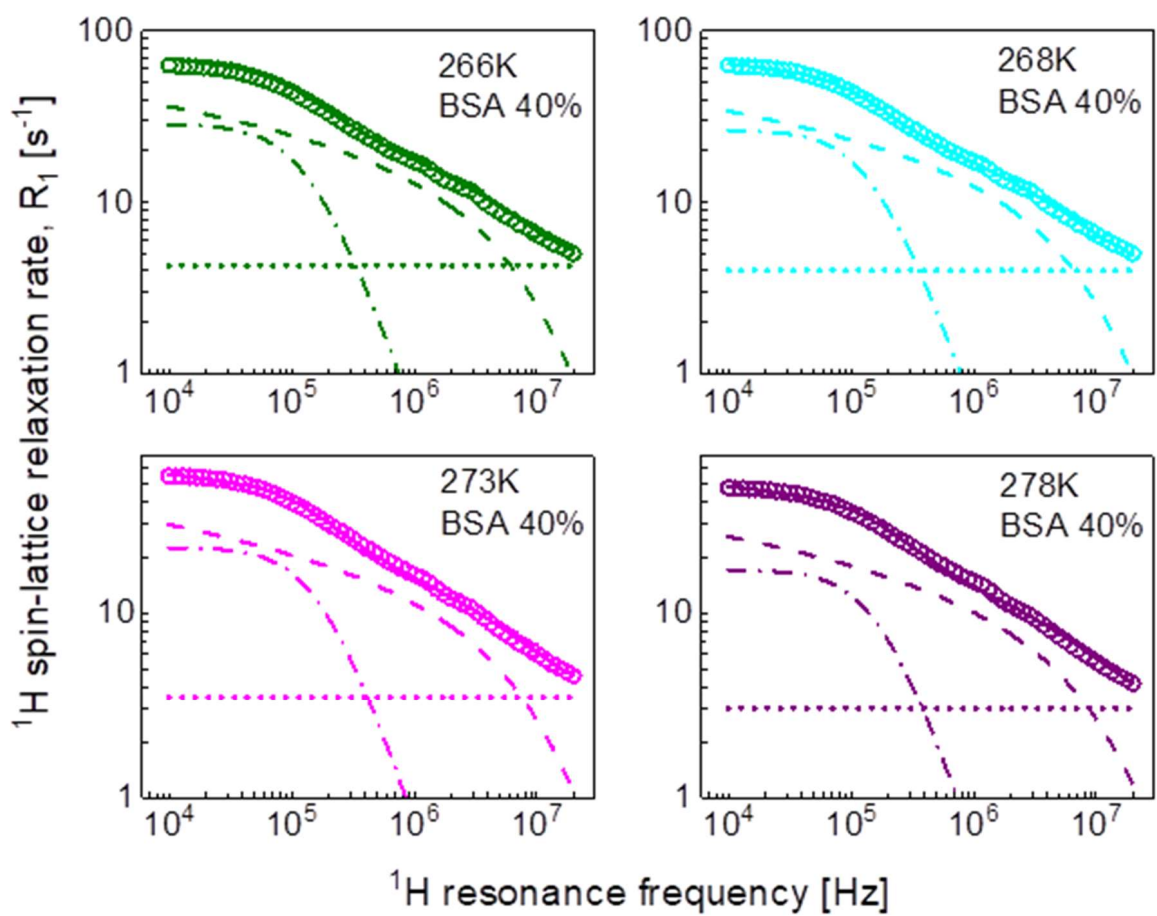

**Figure 6.** $^1$H spin-lattice relaxation data for BSA–water mixture (40%wt of BSA) reproduced in terms of Equation (9); solid line—overall fit decomposed into a Lorentzian term (dashed-dotted line), a term associated with two-dimensional translation diffusion (dashed line) and a frequency independent term (dotted line).

Following the line of two-dimensional translation diffusion, in the last step we have attempted to reproduce the relaxation data in terms of Equation (8) that includes the residence life time, $\tau_{res}$. It has turned out that this concept enables reproducing the relaxation data for 20%wt and 40%wt of BSA with a contribution associated with two-dimensional translation diffusion dominating over a broad frequency range, as shown in Figure 7 and Figure 8, respectively. The obtained parameters are collated in Table 4.

**Table 3.** Parameters obtained from the analysis of the ¹H spin-lattice relaxation data for BSA–water mixtures in terms of Equation (9). The dipolar relaxation constant, $C_{DD}$, for 20% and 40%wt concentration of BSA yield: $C^{DD}=9.81 \times 10^6$ Hz² and $C^{DD}=8.28 \times 10^6$ Hz², respectively, the relaxation constant associated with two dimensional translation diffusion is $C_{trans}=7.09 \times 10^7$ Hz² for both concentrations of BSA. The translation diffusion coefficient has been obtained from the relationship: $D_{trans} = \frac{d^2}{2\tau_{trans}}$.

| Temp. [K] | $\tau_c$ [s] | $\tau_{trans}$ [s] | $A$ [s⁻¹] | $D_{trans}$ [m²/s] |
|---|---|---|---|---|
| | | 20%wt of BSA | | |
| 268 | 1.37 × 10⁻⁷ | 7.26 × 10⁻¹⁰ | 1.70 | 5.02 × 10⁻¹¹ |
| 273 | 1.14 × 10⁻⁷ | 6.15 × 10⁻¹⁰ | 1.55 | 5.93 × 10⁻¹¹ |
| 278 | 9.05 × 10⁻⁸ | 4.05 × 10⁻¹⁰ | 1.19 | 9.00 × 10⁻¹¹ |
| 298 | 5.99 × 10⁻⁸ | 2.50 × 10⁻¹⁰ | 0.77 | 1.46 × 10⁻¹⁰ |
| | | 40%wt of BSA | | |
| 266 | 6.92 × 10⁻⁷ | 7.14 × 10⁻⁹ | 4.32 | 5.11 × 10⁻¹² |
| 268 | 6.39 × 10⁻⁷ | 6.64 × 10⁻⁹ | 4.01 | 5.49 × 10⁻¹² |
| 273 | 5.42 × 10⁻⁷ | 5.67 × 10⁻⁹ | 3.50 | 6.43 × 10⁻¹² |
| 278 | 4.63 × 10⁻⁷ | 4.79 × 10⁻⁹ | 3.03 | 7.61 × 10⁻¹² |

**Table 4.** Parameters obtained from the analysis of the ¹H spin-lattice relaxation data for BSA–water mixtures in terms of Equation (8). The dipolar relaxation constant, $C_{DD}$, for 20% and 40%wt concentration of BSA yield: $C^{DD} = 4.31 \times 10^7$ Hz² and $C^{DD}=9.03 \times 10^7$ Hz², respectively, the relaxation constant associated with two-dimensional translation diffusion is $C_{trans}=8.04 \times 10^6$ Hz² for 20% of BSA and $C_{trans} = 1.07 \times 10^7$ Hz² for 40% of BSA. The translation diffusion coefficient has been obtained from the relationship: $D_{trans} = \frac{d^2}{2\tau_{trans}}$.

| Temp. [K] | $\tau_c$ [s] | $\tau_{trans}$ [s] | $\tau_{res}$ [s] | $A$ [s⁻¹] | $D_{trans}$ [m²/s] |
|---|---|---|---|---|---|
| | | 20%wt of BSA | | | |
| 268 | 6.70 × 10⁻⁹ | 5.17 × 10⁻⁸ | 4.81 × 10⁻⁷ | 2.06 | 7.05 × 10⁻¹³ |
| 273 | 5.46 × 10⁻⁹ | 4.33 × 10⁻⁸ | 4.07 × 10⁻⁷ | 1.90 | 8.42 × 10⁻¹³ |
| 278 | 4.13 × 10⁻⁹ | 3.28 × 10⁻⁸ | 3.12 × 10⁻⁷ | 1.31 | 1.11 × 10⁻¹² |
| 298 | 2.78 × 10⁻⁹ | 2.14 × 10⁻⁸ | 2.09 × 10⁻⁷ | 0.80 | 1.70 × 10⁻¹² |
| | | 40%wt of BSA | | | |
| 266 | 2.38 × 10⁻⁸ | 1.83 × 10⁻⁷ | 2.71 × 10⁻⁶ | 5.06 | 1.99 × 10⁻¹³ |
| 268 | 2.24 × 10⁻⁸ | 1.65 × 10⁻⁷ | 2.66 × 10⁻⁶ | 4.70 | 2.21 × 10⁻¹³ |
| 273 | 1.91 × 10⁻⁸ | 1.33 × 10⁻⁷ | 2.60 × 10⁻⁶ | 4.20 | 2.74 × 10⁻¹³ |
| 278 | 1.66 × 10⁻⁸ | 1.07 × 10⁻⁷ | 2.60 × 10⁻⁶ | 3.51 | 3.41 × 10⁻¹³ |

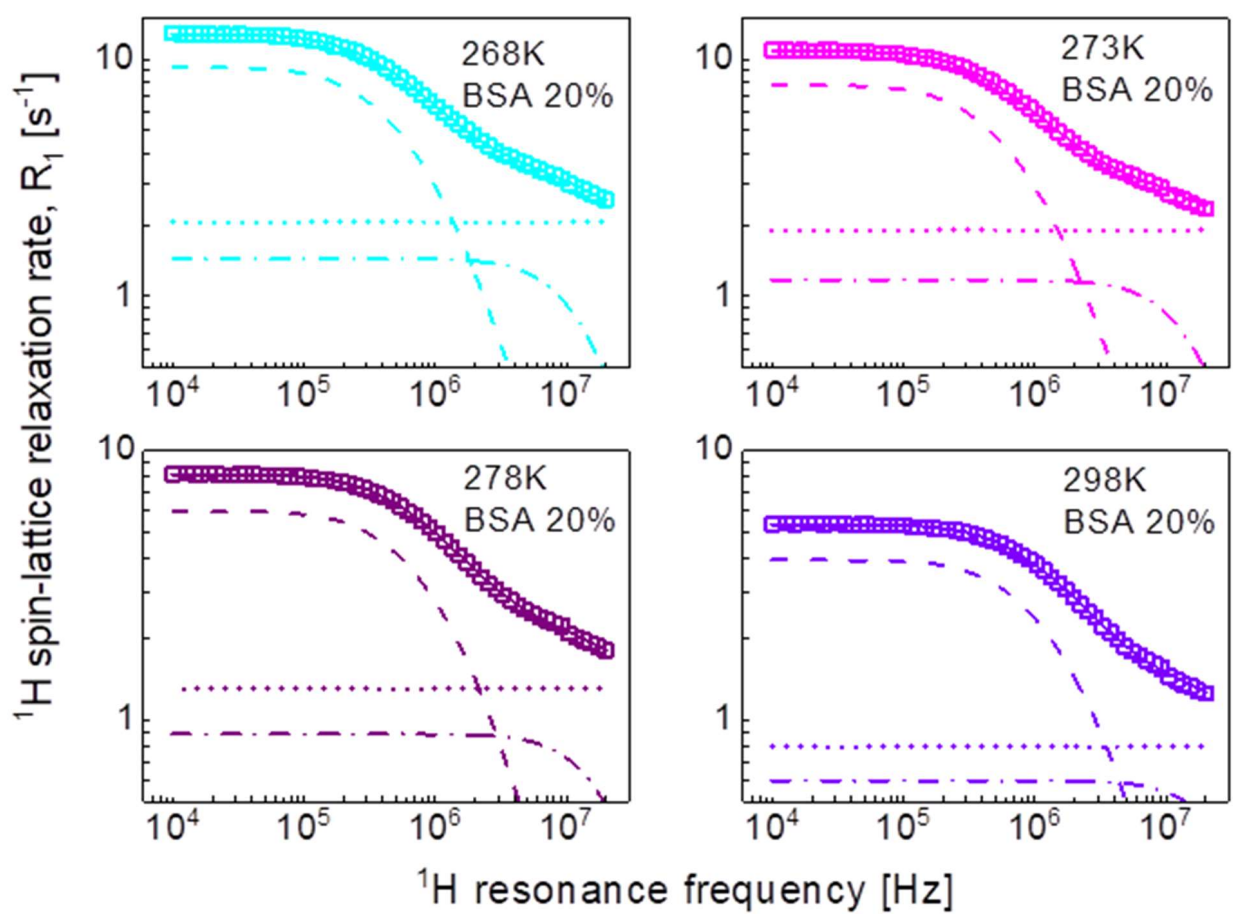

**Figure 7.** $^1$H spin-lattice relaxation data for BSA–water mixture (20%wt of BSA) reproduced in terms of Equation (8); solid line—overall fit decomposed into a term associated with two-dimensional translation diffusion (dashed line), a Lorentzian term (dashed-dotted line) and a frequency independent term (dotted line).

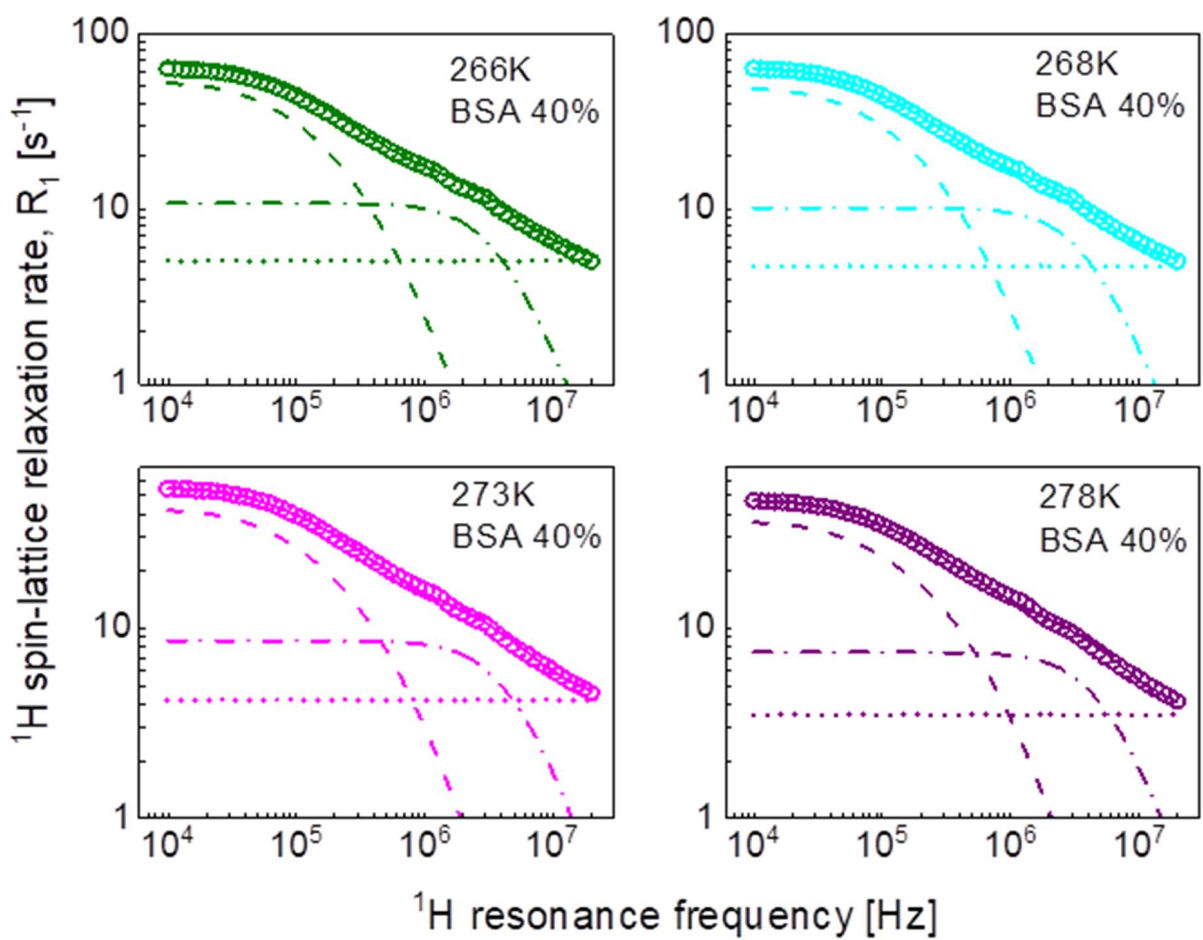

**Figure 8.** $^1$H spin-lattice relaxation data for BSA–water mixture (40%wt of BSA) reproduced in terms of Equation (8); solid line—overall fit decomposed into a term associated with two-dimensional translation diffusion (dashed line), a Lorentzian term (dashed-dotted line) and a frequency independent term (dotted line).

Eventually, we performed a quantitative analysis of the $^1$H spin-lattice relaxation data at lower temperatures, showing QRE effects (quadrupole peaks). The data have been interpreted as a sum of $^1$H-$^1$H and $^1$H-$^{14}$N relaxation contributions. The $^1$H-$^1$H relaxation contribution has been described in terms of Equation (3), while the $^1$H-$^{14}$N relaxation contribution is given by Equation (8). Because of the very similar shape of the relaxation data (Figure 2b)—the data almost overlap after a simple rescaling (multiplication by a factor)—we have limited ourselves to the case of 40%wt of BSA at 263 K (Figure 9). The obtained parameters are as follows: $C_s^{DD}$ =4.92 × 10$^6$ Hz$^2$, $C_i^{DD}$=1.40 × 10$^8$ Hz$^2$, $C_f^{DD}$=5.62 × 10$^8$ Hz$^2$, $\tau_s$=1.31 × 10$^{-6}$ s, $\tau_i$=1.09 × 10$^{-7}$ s, $\tau_f$=1.07 × 10$^{-8}$ s, $C_{DD}^{HN}$=2.27 × 10$^7$ Hz$^2$, $a_Q$ =3.36 × 10$^6$ Hz$^2$, $\eta$ =0.42, $\tau_Q$=1.03 × 10$^{-6}$ s, $\theta$ = 55.4°, $\phi$ = 53.0°, $A$ = 10.8 s$^{-1}$.

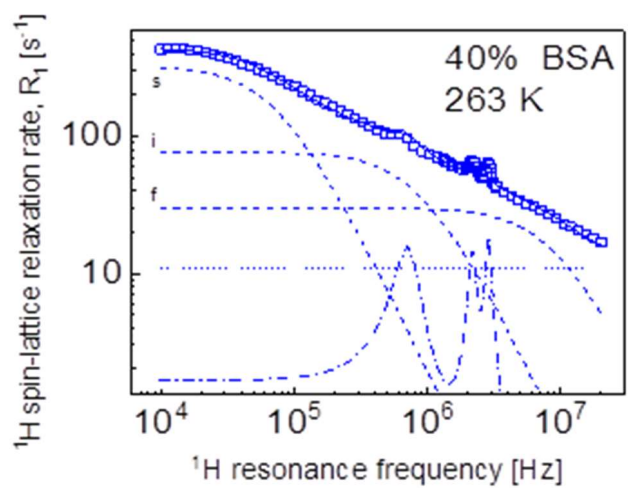

**Figure 9.** $^1$H spin-lattice relaxation data for BSA–water mixture (40%wt of BSA) at 263 K reproduced as a sum of Equation (3) ($^1$H-$^1$H relaxation contribution) and Equation (10) ($^1$H-$^{14}$N relaxation contribution); solid line—overall fit decomposed into the $^1$H-$^1$H relaxation contributions associated with slow, intermediate, and fast dynamics (dashed lines), dashed-dotted line—$^1$H-$^{14}$N relaxation contribution, dotted line—a frequency independent term.

## 3. Discussion

Aiming at revealing the timescale and the mechanism of water motion in highly concentrated water–protein mixtures, in the first step we described the $^1$H spin-lattice relaxation data as a sum of relaxation contributions expressed in terms of Lorentzian spectral densities. The interpretation required two relaxation contributions for the case of 20%wt of BSA and three contributions for the case of 40%wt of BSA. The correlation times are shown in Figure 10. The $\tau_i$ and $\tau_f$ correlation times for 20% and 40%wt BSA are close, the $\tau_s$ values for 40%wt of BSA are by about an order of magnitude longer than $\tau_i$. The correlation times follow (in a good approximation) the Arrhenius law. The dipolar relaxation constant yield: $C_i^{DD}$ = 7.84 × 10$^6$ Hz$^2$, $C_f^{DD}$ = 2.16 × 10$^7$ Hz$^2$ for 20%wt of BSA, $C_s^{DD}$ = 5.61 × 10$^6$ Hz$^2$, $C_i^{DD}$ = 1.45 × 10$^7$ Hz$^2$ and $C_f^{DD}$ = 9.89 × 10$^7$ Hz$^2$ for 40%wt of BSA.

The large number of parameters enables reproducing the relaxation data, although, even then, the agreement is not very good in the whole frequency range. The analysis does not provide indications regarding the mechanism of the dynamical processes associated with the individual relaxation contributions. Therefore, in the next step we have attempted to exploit the model of three-dimensional translation diffusion for water molecules present in the system. However, the relaxation rates at low frequencies do not show linear dependences on $\sqrt{\omega}$. This implies that even water molecules indeed undergo three-dimensional translation diffusion, the relaxation contribution associated with this motion does not dominate the relaxation process in the low-frequency range. This is reflected by the results shown in Figure 3 (20%wt BSA) and Figure 4 (40%wt BSA). Equation (5) used for the relaxation data for 20%wt BSA gives a relaxation contribution associated with three-dimensional translation diffusion that dominates the overall relaxation in a relatively small frequency range, from about 1 MHz to about 5 MHz—at lower frequencies the relaxation contribution expressed in term of Lorentzian spectral densities prevails, while at higher frequencies the frequency independent term takes over. This implies that in this way the concept of three-dimensional translation diffusion can neither be confirmed not excluded, especially as the fits do not show a very good agreement with the data in that range. Nevertheless, Figure 8 includes the correlation times, $\tau_{trans}$, for comparison; the ratio $\tau_{trans}/\tau_f$ is below two. As far as the dipolar relaxation constant, $C^{DD}$, and the correlation time, $\tau_c$, are concerned, the quantities are very close to $C_i^{DD}$ and $\tau_i$. The obvious gain from using the model of three-dimensional translation diffusion for the case of 40%wt of BSA is reducing the number of parameters. The relaxation contributions expressed in terms of $C_i^{DD}$, $\tau_i$ and $C_f^{DD}$, $\tau_f$ have been replaced by a relaxation contribution including only two adjustable parameters: $N_H$ and $\tau_{trans}$, leading to a better (not worse) agreement with the experimental data (Figure 4). The parameters $C^{DD}$ and $\tau_c$ do not differ considerably from $C_s^{DD}$ and $\tau_s$. The values of the correlation time, $\tau_{trans}$, are shown in Figure 8, the ratio $\tau_i/\tau_{trans}$ is of about two.

This scenario of motion, although plausible at the first sight (not discussing at this stage the origin of the dynamics reflected by the relaxation contribution with Lorentzian spectral densities, characterized by $C^{DD}$ and $\tau_c$) has turned out to be unambiguous. The relaxation data for both 20%wt and 40%wt BSA can also be reproduced in terms of Equation (9), in which the relaxation contribution associated with three-dimensional translation diffusion has been replaced by a relaxation contribution corresponding to two-dimensional translation diffusion under the assumption $\frac{\tau_{trans}}{\tau_{res}} < \omega \tau_{trans}$ (long $\tau_{res}$). The success of this undertaking is not surprising for 20%wt of BSA as in this case the relaxation is anyway dominated by the relaxation term including Lorentzian spectral densities. However, the agreement with the experimental data reached for 40%wt of BSA renders the mechanism of the translation diffusion unambiguous. It is worth to mention at this stage the short correlation times obtained for the two-dimensional translation diffusion compared to those for three-dimensional motion.

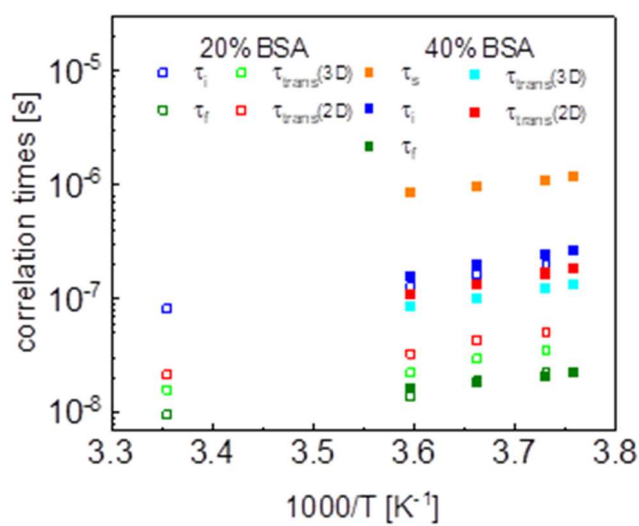

**Figure 10.** Comparison of correlation times obtained by means of different models.

This discussion brings one to the conclusion that to resolve the question about the mechanism of molecular motion it is required that the relaxation contribution associated with the dynamical process dominates the overall relaxation over a relatively broad range of resonance frequencies—otherwise the characteristic features of the spectral density functions can be masked by other relaxation contributions. This has been achieved for the model given by Equation (8). The relaxation terms associated with two-dimensional translation diffusion gives a dominating contribution over at least two decades of frequency. The correlation times characterizing the two-dimensional translation motion for 20%wt and 40%wt BSA are included in Figure 10. They show that the translation diffusion of water molecules in the 40%wt BSA mixture is about three times slower compared to the diffusion in the 20%wt BSA mixture. For the case of 20%wt BSA, the residence lifetime of water molecules on the protein surface is by an order of magnitude longer than the correlation time for the translation diffusion, while for the case of 40%wt of BSA the ratio is about five, the residence lifetime being weakly temperature dependent (almost independent). The model of Equation (8) also includes a relaxation contribution represented in terms of Lorentzian spectral densities. The correlation time, $\tau_c$ for the 40%wt of BSA, is close to the $\tau_f$ values obtained from the analysis in terms of relaxation contributions with Lorentzian spectral densities (Equation (3)). For the case of 20%wt of BSA, the correlation time $\tau_c$ is shorter than $\tau_f$.

On the basis of the performed analysis one can construct the following scenario of the water dynamics. There is a fraction of water molecules undergoing two-dimensional translation diffusion in the vicinity of the protein surface. The diffusion is interrupted by adsorption on the protein surface for a time characterized by $\tau_{res}$. The absorbed water molecules follow the rotational dynamics of the protein molecules, characterized by the correlation time $\tau_c$. This concept is supported by the values of the dipolar relaxation constants, $C^{DD}$, obtained from the model of Equation (9), they yield: 4.31 × 10$^7$ Hz$^2$ for 20%wt of BSA and 9.03 × 10$^7$ Hz$^2$ for 40%wt of BSA. The dipolar relaxation constant is proportional to the mole fraction of bound water molecules [44]—the increase of the BSA concentration by factor two leads to as similar increase in the fraction of bound water molecules. The frequency independent term, $A$, corresponds to a fraction of water molecules the dynamics of which is affected by interactions with the macromolecules to a much lesser extend—the dynamics remain so fast that there is no frequency dependence of this relaxation contribution in the covered frequency range.

Eventually it is worth comparing the parameters obtained from the analysis of the relaxation data for 40%wt BSA at 263 K (Figure 9) with those obtained for solid BSA [16]. The quadrupole parameters, $a_Q$, $\eta$ and $\tau_Q$ are almost the same as expected—they describe the properties of the electric field gradient tensor at the position of $^{14}$N nuclei in the protein backbones. The correlation times $\tau_s$ and $\tau_i$ are shorter compared to the case of the solid protein by about factor two, while $\tau_f$ remains the same. Performing this comparison one should keep in mind that the model of Equation (3) should be treated as only a parametrization of the data. For instance, the dynamical process referred to as the slow dynamics ($\tau_s$) for 40%wt BSA is not matched by a process occurring on a similar time scale when the models involving translation diffusion are applied.

Finishing the discussion, we wish to point out that other NMR methods (NMR spectroscopy and diffusometry) are highly appreciated as a source of information about protein (biomolecular) systems [45–48].

## 4. Materials and Methods

Bovine serum albumin (BSA) lyophilized powder was bought from Merck® company. Both BSA solutions were prepared by dissolving 0.25g (20% concentration) and 0.67g (40% concentration) of BSA powder in 1ml of PBS (Phosphate buffered saline) at room temperature with slow stirring (250 RPM) on the magnetic stirrer. Two mg (one tablet) of solid PBS was dissolved in 200 mL of deionized water; pH was 7.4 at 25 °C. BSA powder was added to 1 mL of PBS in portions over a period of 5 h and then transferred into 10 mm diameter NMR tube. After preparing, the solutions were stored in the fridge.

$^1$H spin-lattice relaxation measurements have been performed in the frequency range from 10 kHz to 20 MHz versus temperature from 298 K to 253 K, using a "1 Tesla NMR relaxometer", produced by Stelar s.r.l. (Mede (PV), Italy). The temperature was controlled with an accuracy of 0.5 K using a built-in VTC temperature controller. For measurements performed at 298 K and above, the carrier gas was dry, compressed air, while for temperatures below 298 K, was nitrogen. The experiments started from 298 K and then the temperature was progressively decreased, up to 253 K. The switching time of the magnet was set to 3 ms. The pre-polarization was applied below 10 MHz. For all temperatures, 60 values of T$_1$ (R$_1$ = 1/T$_1$) in the whole frequency range were acquired. Additionally, for those profiles where QRE peaks appeared, 40 more values of T$_1$ were collected in 3.3–1.8 MHz range. For each resonance frequency, 32 magnetization values have been recorded versus

time in a logarithmic time scale. The relaxation processes have turned out to be single-exponential for all temperatures in the whole frequency range for both concentrations. Examples of the magnetization curves ($^1$H magnetization versus time) are shown in the Appendix A.

## 5. Conclusions

$^1$H spin-lattice relaxation studies have been performed for BSA–water mixtures (20%wt of BSA and 40%wt of BSA) in the frequency range from 10 kHz to 10 MHz, versus temperature. The data have been used to enquire into the mechanism of water motion. For this purpose, four models have been applied. In the first step, the data have been parametrized in terms of relaxation contributions expressed by Lorentzian spectral densities. The large number of parameters has allowed to reproduce the relaxation data—one should note that in the case of 20%wt of BSA only two relaxation contributions are needed, while in the case of 40%wt of BSA requires three relaxation contributions. In the next step, one of the Lorentzian terms has been replaced by the model of three-dimensional translation diffusion. At the first sight the concept has turned out to be successful, however a closer inspection of the decomposition of the overall relaxation rates raised doubts regarding unambiguity of the analysis. The relaxation contribution supposedly associated with the translation diffusion does not dominate the relaxation process in the low frequency range. Consequently, one cannot profit from the mathematical features characteristic of the spectral density associated with three-dimensional translation diffusion. In the third step, the model of three-dimensional translation diffusion has been replaced by a relaxation term assuming two-dimensional translation motion. In this case, due to a significant contribution of a relaxation contribution expressed in terms of Lorentzian spectral densities in the low frequency range, the mathematical properties of the corresponding spectral density functions could not be used as a discriminating factor. These examples demonstrate that unambiguous analysis of NMR relaxometry data for complex molecular systems requires situations in which the relaxation data follow the mathematical form of a specific spectral density over a broad frequency range and this effect is not masked by other relaxation contributions. This has been achieved for the model of two-dimensional translation diffusion modulated by acts of adsorption to the surface with a residence lifetime not being much longer (orders of magnitude) than the correlation time of the translation motion, rendering the conclusion that the diffusion process is of two-dimensional character. The presented strategy of the data analysis demonstrates the need for thorough evaluation of the applied models to profit from the potential of NMR relaxometry.

**Author Contributions:** Conceptualization, D.K.; methodology, D.K. and E.M.; software, R.C. and B.N.; investigation, A.K., E.M., and K.K.; writing—original draft preparation, D.K.; writing—review and editing, D.K. and E.M.; funding acquisition, D.K. All authors have read and agreed to the published version of the manuscript.

**Funding:** This research has received funding from the European Union's Horizon 2020 research and innovation program under grant agreement No 899683 (project "HIRES-MULTIDYN").

**Institutional Review Board Statement:** Not applicable.

**Informed Consent Statement:** Not applicable.

**Data Availability Statement:** https://doi.org/10.5281/zenodo.7595111

**Conflicts of Interest:** The authors declare no conflict of interest.

## Appendix A

Selected $^1$H magnetization curves ($^1$H magnetization versus time) for BSA–water mixtures.

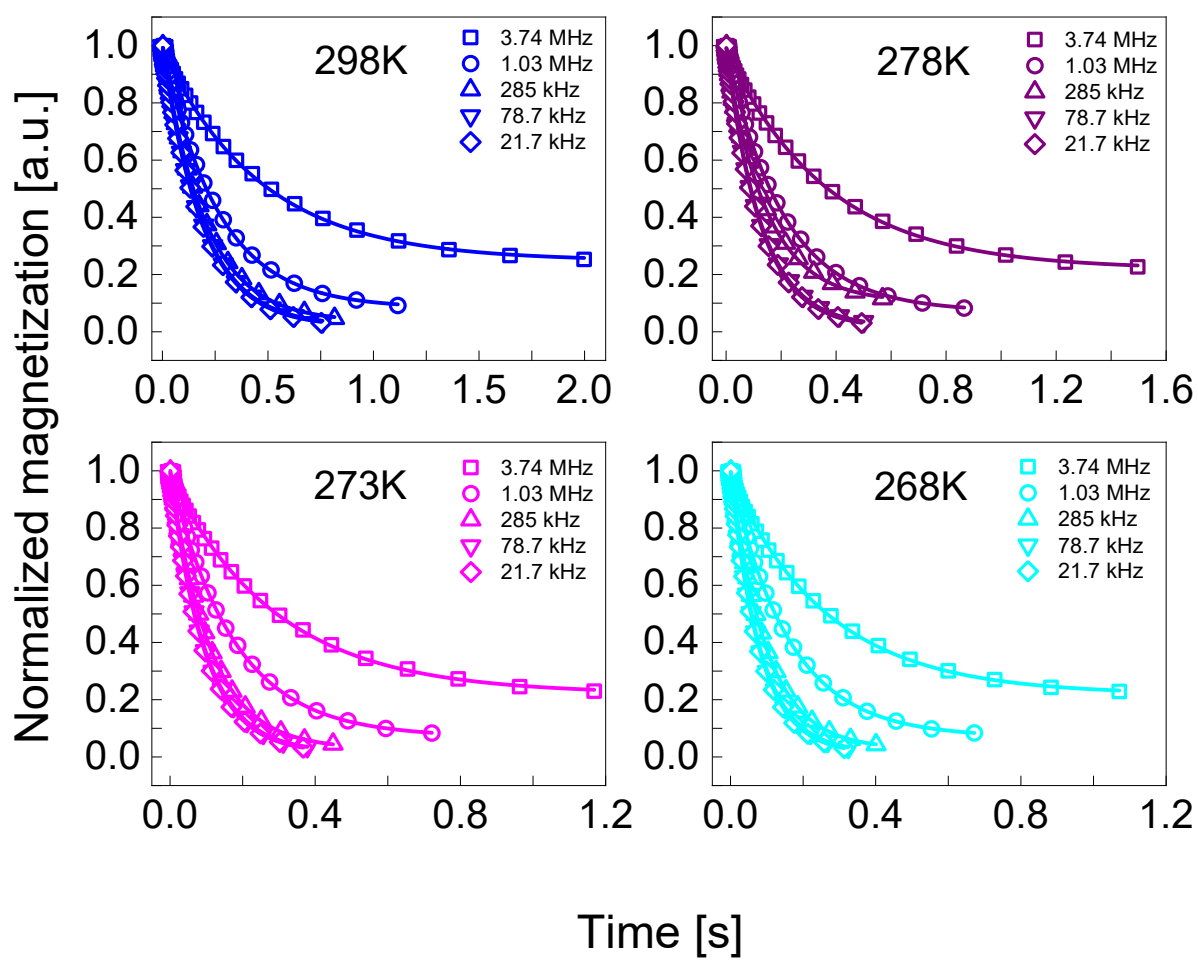

**Figure A1.** Normalized ¹H magnetization curves for BSA–water mixture (20%wt BSA) at selected resonance frequencies in the temperature range from 298 K to 268 K. Solid lines—single exponential fits.

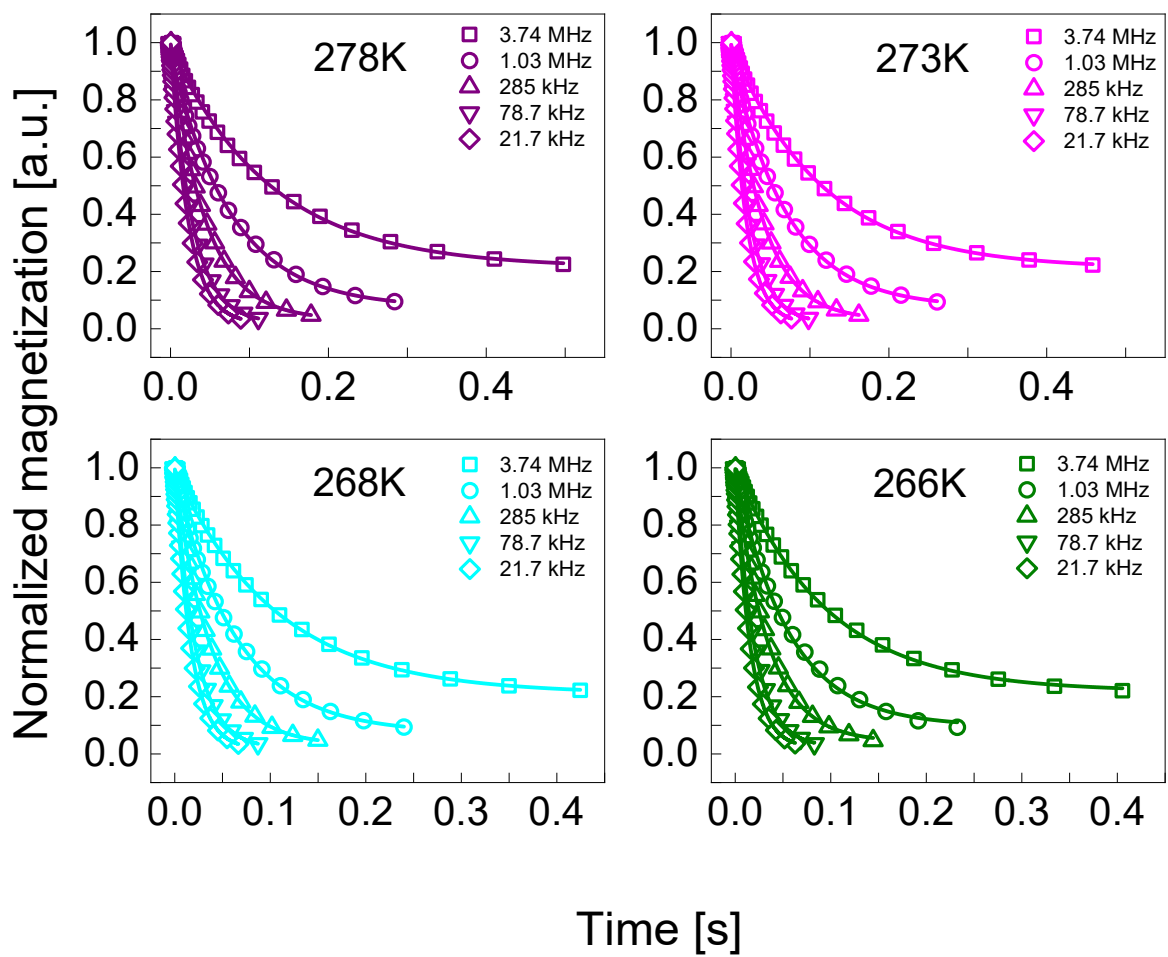

**Figure A2.** Normalized ¹H magnetization curves for BSA–water mixture (40%wt BSA) at selected resonance frequencies in the temperature range from 278 K to 266 K. Solid lines—single exponential fits.

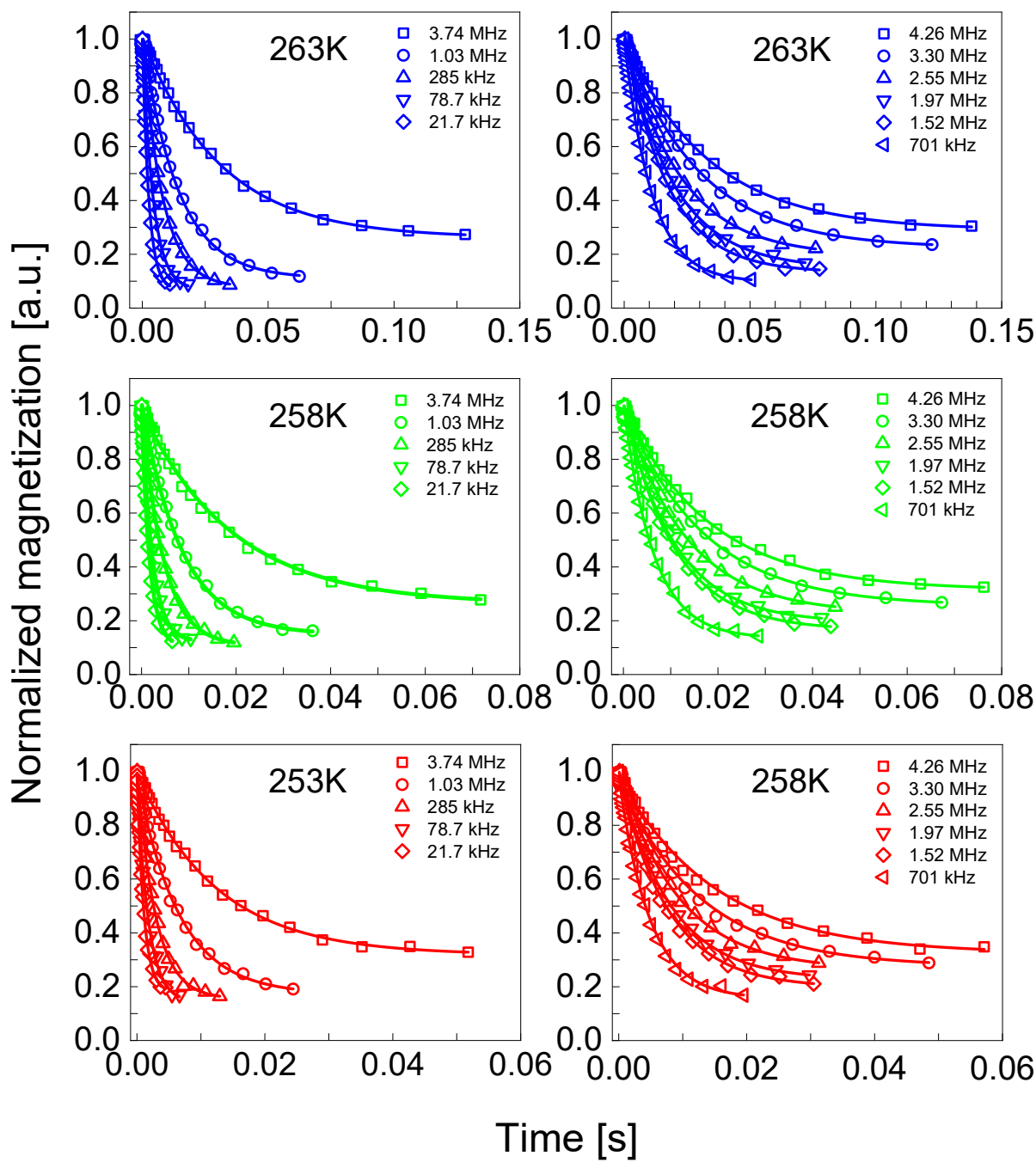

**Figure A3.** Normalized ¹H magnetization curves for BSA–water mixture (20%wt BSA) at selected resonance frequencies in the temperature range from 263 K to 253 K. Solid lines—single exponential fits. On the left frequencies are selected from the entire range measured and, on the right, only from areas with quadrupole peaks.

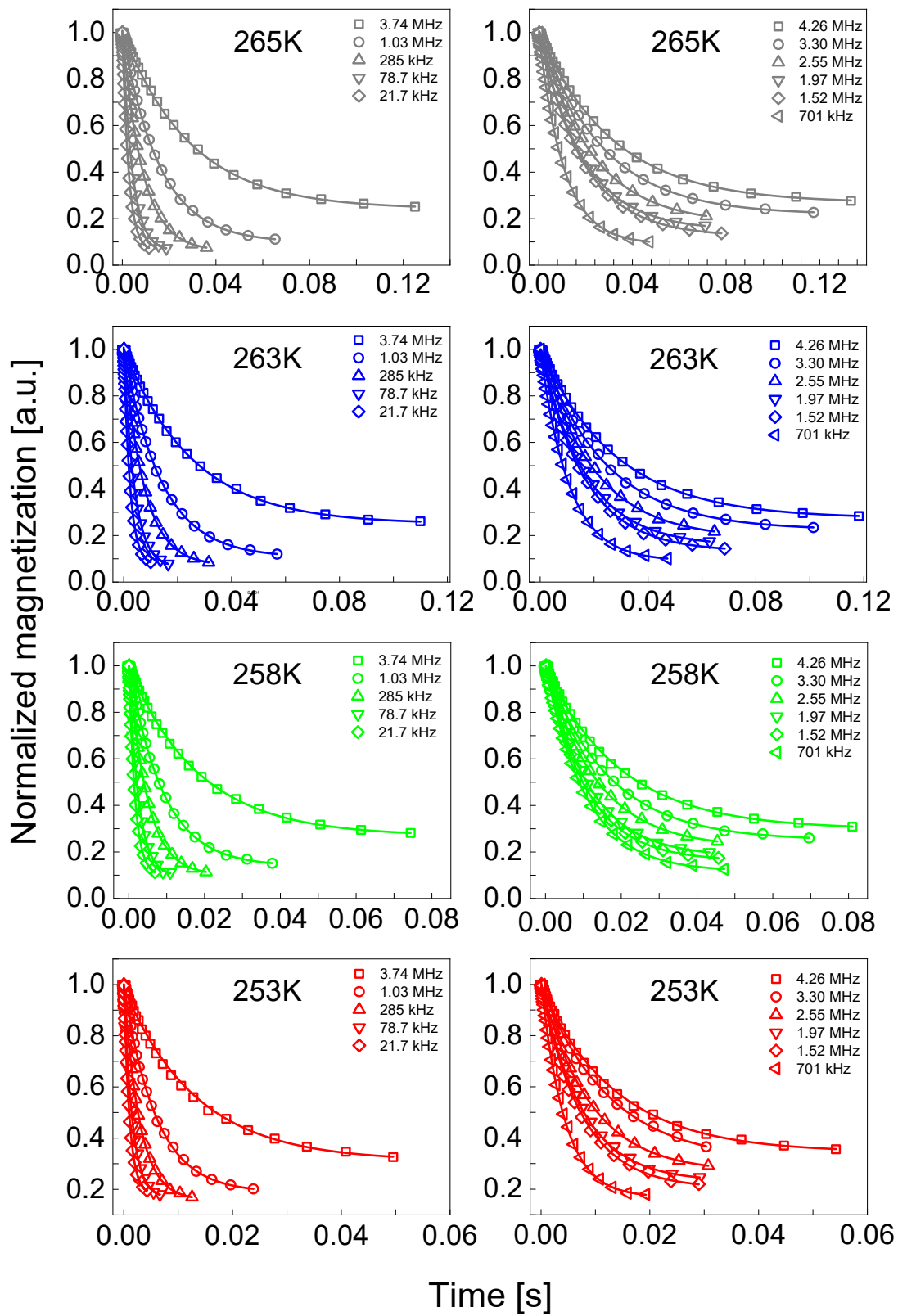

**Figure A4.** Normalized ¹H magnetization curves for BSA–water mixture (40%wt BSA) at selected resonance frequencies in the temperature range from 263 K to 253 K. Solid lines—single exponential fits. The frequencies are selected from the entire range measured.